# TRANSIENT STRUCTURED FLUCTUATIONS IN A TWO-DIMENSIONAL SYSTEM WITH MULTPLE ORDERED PHASES

*Zach Krebs[†], Ari B. Roitman[†], Linsey M. Nowack[†], Chris Liepold[†], Binhua Lin[†‡]\*, and Stuart A. Rice[†§]\**

[†] James Franck Institute, University of Chicago, Chicago, IL 60637, USA

[‡] Center for Advanced Radiation Sources, University of Chicago, Chicago, IL 60637, USA

[§] Department of Chemistry, University of Chicago, Chicago, IL 60637, USA

Corresponding Author:  sarice@uchicago.edu

**Abstract**

We report the structure of transient fluctuations in the liquid phase of a two-dimensional system that exhibits several ordered phases with different symmetries.  The density-temperature phase diagram of the system studied, composed of particles with a repulsive shouldered soft-core pair interaction, has regions with stable liquid and hexatic phases, a square solid phase, two separate hexagonal solid phases and a quasi-crystalline phase with 12-fold symmetry.  We have examined the character of the structured fluctuations by computing the same-time aperture cross correlation function of particle configurations in several fluid regions near to and far from phase transition lines.  The primary goals of our study are: (1) Determination if the spectrum of structures of the fluctuations in the liquid is broader than or limited to the motifs exhibited by the ordered phases supported by the system, and (2) Determination of the density domains in the liquid that support particular transient structured fluctuations.  In the system studied, along a low-temperature isotherm in the temperature-density plane that intersects all the ordered phases we find that the liquid phase exhibits structured fluctuations with hexagonal symmetry near both liquid-hexatic transition lines.  Along the same isotherm and in the stable liquid between the lower density hexatic-to-liquid and the higher density liquid-to-square solid transitions, we find that transient hexagonal ordered fluctuations dominate the liquid region near the hexatic-to-liquid transition and square ordered fluctuations dominate the liquid region near the liquid-to-square solid transition, but that both structured fluctuations occur at all densities between these transition lines.  At a higher temperature, at phase points in the liquid above but close to the density maximum of an underlying transition, there are ordered fluctuations that can be correlated with the structure of the lower temperature phase.  Although it is expected that very close to a liquid-ordered phase boundary a structured fluctuation in the liquid will have the same symmetry as the ordered phase, it is not obvious that structured fluctuations in thermodynamic states deep in the liquid phase will be similarly restricted. The most striking result of our calculations is that no evidence is found in the liquid phase for structured fluctuations with other symmetries than those of the ordered phases of the system.

## 1. Introduction

Until very recently it was believed that in a two-dimensional (2D) system composed of particles that interact via a short-ranged central pair potential, the most stable packing of the particles is in a hexagonal lattice, based on the assertion that the lowest energy configuration of particles that interact via central forces has the maximum number of nearest neighbours. However, recent research has revealed that there exist simple monotone repulsive central pair potentials that support, in a 2D system, stable ground state (zero temperature) lattices with hexagonal packing, square packing, Kagome packing, and many other packing motifs [1-7], and that these lattices have vibrational spectra with only positive frequencies hence are stable over non-trivial density and temperature ranges. For a system with monotone repulsive pair potential the free energy per particle, hence the stability of a particular lattice relative to other lattices, is determined by delicate differences between the values of both the pair potential and the pair force at the locations of the first few coordination shells of the several lattice structures [4]. In a dense liquid the distribution of the number of neighbours in the domain up to the first minimum in the ensemble averaged pair correlation function has a standard deviation that implies the existence of many instantaneous configurations with first shell coordination numbers considerably different from the ensemble averaged value [8,9]. Then, in view of the delicate balance that results in the selection of a particular packing motif for the ordered phase out of a set of different packing motifs, it is plausible to expect that in the liquid phase there will be structured fluctuations representing both the most stable packing motif and other less stable motifs, albeit with different probabilities.

This paper examines the structures, specifically the local particle orderings, of transient fluctuations in the liquid phase of a 2D system whose phase diagram exhibits several ordered phases with different symmetries. The primary goals of our study are: (1) Determination if the spectrum of structures of the fluctuations in the liquid is broader than or limited to the motifs exhibited by the ordered phases supported by the system, and (2) Determination of the density domains in the liquid that support particular transient structured fluctuations. We examine a monitor of the fluctuations in the system that can be determined from experimental measurements and provides a unique signature of the structure of a fluctuation. That monitor is the aperture cross correlation function (ACCF), obtained from simultaneous measurement at two different values of momentum transfer of the intensities of coherent radiation scattered from a domain in the liquid with linear dimension comparable to the correlation length [10]. Although this paper focuses attention on transient structured fluctuations in a particular 2D liquid generated by computer simulation, a subsidiary goal of the research reported is promotion of the use of the ACCF in studies of disordered 2D and 3D dense matter.

Fluctuations of the properties of the liquid phase have been studied for more than a century. For present purposes, these studies can be crudely divided into two broad categories. The first category considers fluctuations as small deviations from the mean values of thermodynamic variables and calculates the probability distributions of these deviations [11]. This analysis does not ascribe any structure to fluctuations; the properties of the fluctuations are related to the extensive and intensive properties of the liquid by standard thermodynamic relations—e.g. the fluctuation in density depends on the temperature and isothermal compressibility. This version of the theory of fluctuations is the canonical basis for the interpretation of light scattering from a liquid or a liquid solution. Using the representation of thermodynamic functions as integrals over combinations of molecular distribution functions, this theory is also the basis for the Kirkwood-Buff theory of solutions [12].

In the second category, fluctuations are identified with local ordered particle configurations in a liquid, but this quality of being "ordered" is defined by an arbitrary proximity to selected symmetric structures [13-15]. For example, if the local inter-particle separations and "bond" angles around a selected particle in a three-dimensional liquid are different from perfect dodecahedral values by less than arbitrary predetermined amounts, dodecahedral symmetry is assigned to the surroundings of the particle. This approach has been used to search particle configurations obtained from Monte Carlo and Molecular Dynamics simulation data for unsuspected local ordering in a liquid and to inform the analysis of nucleation of an ordered phase [16]. The weaknesses of this analysis are its dependence on an arbitrary definition of proximity to a selected ordered structure and the lack of a well-defined experimental procedure from which can be obtained a unique signature of the selected structure.

The conventional characterizations of local order in a liquid are usually not directly testable with real world experiments. The most common characterization of the structure of a liquid, namely the pair correlation function, does not provide a signature of the angular symmetry of local ordering in the liquid; it describes only the angle averaged probability of finding another particle a distance $r$ from a selected particle. The pair correlation function is derived from a conventional diffraction experiment that provides an isotropic average over all the local structural motifs in the illuminated volume, thereby generating a one-dimensional intensity versus angle distribution. However, characterization of the local order around a particle requires information about the angular distribution and locations of many particles, typically the three- and four-body correlation functions. In contrast, angular correlations between the intensities in the two-dimensional transmission diffraction from a limited volume do contain information about the local structure in that volume; it can be extracted via cross correlation of the simultaneous intensities of the scattered radiation at different angles [10]. Thus, if scattering from an area (volume) about the size of the correlation area (volume) in the 2D (3D) liquid is measured simultaneously with two detectors with variable angular separation, the cross correlation of the intensities will have peaks at the angles corresponding to the Bragg scattering peaks of the local structure, if any.–The ACCF is one of a category of cross-correlation fluctuation diffraction descriptors that provide unambiguous signatures of the symmetry of local order in a disordered system. Theoretical descriptions of the experimental method and some applications can be found in Ref. [16-22]

We now confine our focus to the properties of a 2D system whose phase diagram exhibits several ordered phases with different symmetries. We expect the occurrence and nature of the transient structured fluctuations in the liquid state of this system to depend on distance from the liquid-ordered phase boundaries. The location of phase boundaries and the character of phase transitions in 2D systems has been the subject of numerous studies for the last eighty years [17,23-26]. The key observations are that (i) a 2D system cannot support a fully ordered crystalline phase with long-range non-decaying translation order but can support a partially ordered solid phase with algebraically decaying translation order and long range "bond" orientation order, and (ii) as a consequence it is possible for a 2D system to support a phase intermediate between the disordered liquid and the partially ordered solid that has short-range translation order and algebraically decaying orientation order, a so-called hexatic phase [27]. The existence of the hexatic phase was predicted by the Kosterlitz, Thouless, Halperin, Nelson and Young (KTHNY) theory of 2D melting [27-32]. This theory characterizes the 2D solid as a continuous deformable medium with inclusion of the two classes of point topological defects that have smallest excitation energy to mediate structural changes; it relates the melting process to the mechanical instability of the 2D solid. The theory makes specific predictions concerning the rates of decay of the envelopes of the pair correlation function and the local bond orientation function (see Section 3). The former correlation function is measurable via diffraction of radiation from the liquid. The latter correlation function is measurable only in exceptional cases that permit recording of images of all of the particles; it usually must be calculated from particle configurations generated by computer simulations. The values of the correlation function decay rates at the solid-to-hexatic and the hexatic-to-liquid transition densities are used to locate the respective transition densities. We will make reference to particular predictions obtained from the KTHNY theory at several points in the following text, but our study of structured fluctuations is not related to KTHNY theory per se. Indeed, because it is based on the representation of the system free energy using continuum elastic constants, the KTHNY Hamiltonian does not account for steric effects associated with the non-zero size of the particles in the system. Although the bond orientation function is a measure of local order, and the decay of its envelope is defined by the elastic interactions described with the KTHNY Hamiltonian, that function must be calculated from particle configurations that are not obtained from the KTHNY Hamiltonian. The roles played by excluded volume and local structure are subordinated to representation by the core energy of a dislocation. Consequently, the existence of transient ordered fluctuations in the 2D liquid phase is not addressed and the ACCF of the 2D liquid provides information not obtainable from the KTHNY analysis.

  A 2D system whose phase diagram exhibits several ordered phases with different symmetries can be supported by a pair interaction with a shoulder on an otherwise decaying repulsion.[27,28] For the studies reported in this paper we have used a pair potential introduced by Ryzhov and coworkers [35-37]; that potential is (see Fig. 1)

$$u(r) = \varepsilon \left(\frac{\sigma}{r}\right)^{14} + \frac{1}{2}\varepsilon\{1 - tanh(k[r - \sigma_1])\}. \qquad (1.1)$$

In (1.1) $r$ is the particle separation, $\varepsilon$ the depth of the potential well, $k$, $\sigma$ and $\sigma_1$ are, respectively, an inverse scale length and length scales that characterize the interaction. We will use the value $k = 10/\sigma$ for all the calculations reported in this paper. Ryzhov and coworkers have shown that the density-temperature ($\varrho,T$) phase diagram (Fig.2) of a 2D system with this pair interaction has two domains in which a hexagonal (or triangular solid, T) phase is stable, corresponding to the two characteristic lengths $\sigma$ and $\sigma_1$, and a domain in which a square solid (S) is stable [35,36]. The phase diagram also has regions in which, respectively, liquid and hexatic phases are stable. We note that in a later publication [37] Ryzhov and coworkers have shown that the potential (1.1) also supports a quasi-crystalline phase with 12-fold symmetry. At low temperature the quasi-crystalline phase occupies a very narrow domain between the square and triangular phases. The quasi-crystalline phase will not be further considered in this paper.

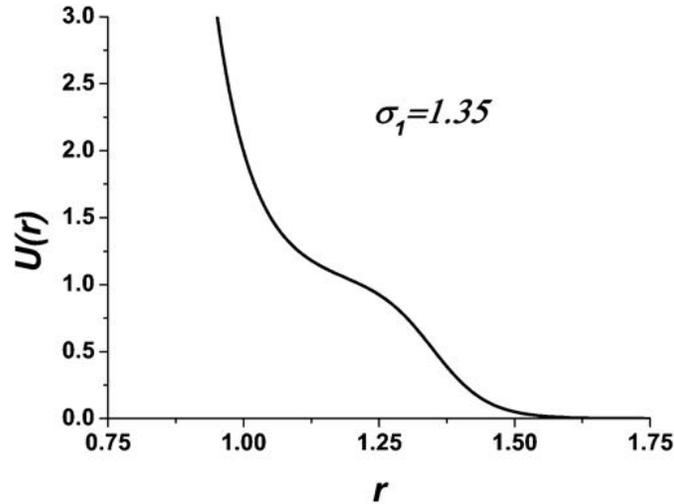

**Figure 1.** The model 2D system pair potential (Eq. (1.1)). Reproduced from *Soft Matter* **10**, 4966 (2014) with permission from The Royal Society of Chemistry.

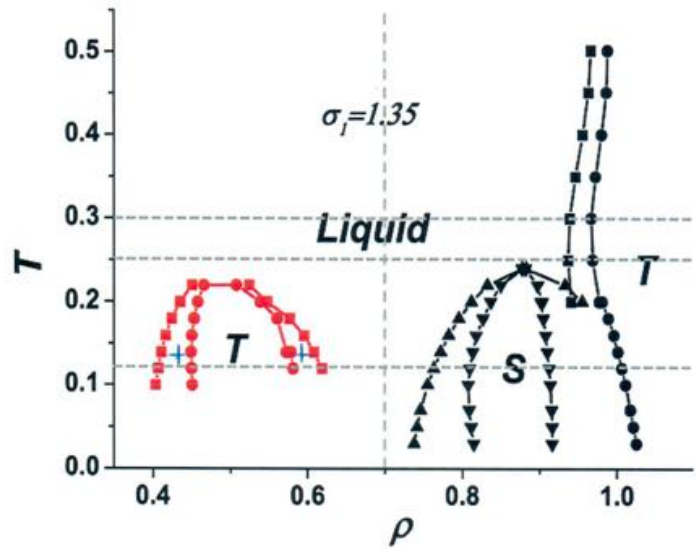

**Figure 2.** The phase diagram of the 2D system with pair potential given by Equation (1.1). Reproduced from *Soft Matter* **10**, 4966 (2014) with permission from the Royal Society of Chemistry. The horizontal dashed lines indicate isotherms $T = 0.12$, $0.25$ and $0.3$ and the vertical dashed line indicates the isochore $\rho = 0.70$, along which simulation were carried out. The pluses (+) identify the hexatic phases.

Following the dashed lines in Fig. 2, along a single low temperature isotherm it is possible, starting at low density and proceeding to high density, to watch the system undergo liquid-to-hexatic, liquid-to-triangular-solid, triangular solid-to-hexatic, hexatic-to-liquid, liquid-to-square-solid and (omitting consideration of the quasi-crystalline phase that is not shown) square solid-to-triangular solid transitions. The hexatic-to-solid and hexatic-to-liquid transitions on the low-density side of the phase diagram are continuous, whereas the transitions on the high-density side of the phase diagram are first order. Returning to Fig. 2, along a high temperature isotherm (topmost dashed line) the system undergoes only one liquid-to-triangular-solid transition. And it is possible to pick a line of constant density (vertical dashed line) that lies entirely in the liquid phase. Our expectations for the structures of transiently ordered fluctuations in different regions of the phase diagram of a system with pair potential given by Equation (1.1) are guided by the results of Sheu and Rice's study of the structured transient fluctuations in the quasi-two-dimensional (q2D) near hard sphere fluid confined between smooth hard walls [38]. This system exhibits alternating hexagonal and square packed ordered solids as the wall separation is increased from one hard sphere diameter to 1.57 diameters, to 1.75 diameters, and on [39]. Their results reveal that transient fluctuations in the q2D fluid have a limited range of structural motifs and that the precursor fluctuations to a specific ordered structure do not all exhibit the same structure as the ordered solid.

With respect to the 2D system with pair potential (1.1), our simulation results reveal that at low temperature, on the low and the high-density sides of the transition to the hexatic phase that lies between the liquid and the triangular solid phase, there are structured fluctuations with hexagonal symmetry in the liquid. There is no evidence for ordered fluctuations in the liquid with any other symmetry. Along the same isotherm, at higher density the system undergoes a first order transition to a square solid phase. In the stable liquid between the lower density hexatic-to-liquid and the higher density liquid-to-square solid transitions we find that transient hexagonal ordered fluctuations dominate the region near the hexatic-to-liquid transition and square ordered fluctuations dominate the region near the liquid-to square solid transition, but that both structured fluctuations occur at all densities between these transition lines. At higher temperatures, at phase points in the liquid above but close to the density maximum of an underlying transition, there are ordered fluctuations that can be correlated with the structure of the lower temperature phase. Although it is to be expected that very close to a liquid-ordered phase boundary a structured fluctuation in the liquid will have the same symmetry as the ordered phase, it is not obvious that structured fluctuations in thermodynamic states deep in the liquid phase will be similarly restricted. The most striking result of our calculations is that no evidence is found in the liquid phase for structured fluctuations with other symmetries than those of the ordered phases of the system.

## 2. Model System and Calculations

We have carried out constant temperature and volume molecular dynamics simulations of 2D systems of 8700 particles that interact with the pair potential displayed in Equation (1.1) using the 2D version of the LAMMPS package [40]. The results obtained from sample simulations with larger numbers of particles (20,000) were negligibly different from those with 8700 particles. The phase boundaries determined from our simulations duplicate those reported by Ryzhov and co-workers within the precision expected from the use of different numbers of particles in the simulations.

Simulation runs that sample most of the phase diagram were typically $10^6$ time-steps long, with each time-step of duration $0.001\ \tau$, where $\tau \equiv \sqrt{m\sigma^2/\varepsilon}$. The $10^6$ time-steps of each simulation run were divided into 10,000 "frames", with a frame representing the output of particle coordinates and velocities every 100 time-steps. To determine that the system had reached equilibrium in the last 1,000 frames of the simulation, the total energy of the system was monitored every 1,000 time-steps. We found that at low temperature and high density $10^6$ time-steps were not enough to reach equilibrium. To drive these systems to equilibrium a perturbation that generates pathways out of metastable states, specifically a simulation cell deformation, was applied. We describe the results of our simulations using the conventional reduced variables $r^* \equiv \frac{r}{\sigma}, P^* \equiv \frac{P\sigma^2}{\varepsilon}, V^* \equiv \frac{V}{N\sigma^2} \equiv \frac{1}{\rho^*}$, and $T^* \equiv \frac{k_B T}{\varepsilon}$. Since only reduced variables are referred to hereafter, the asterisks on these variables will not be displayed in the remainder of this paper.

The particle configurations generated by the simulations were analyzed via their structure functions and aperture cross correlation functions. Here, we imagine that an incident plane wave with wave vector $\boldsymbol{k}_i$ illuminates a small region containing $N$ particles, defined by an aperture. The linear dimensions of the aperture are comparable with the correlation length in the system. The scattered wave that emerges has wave vector $\boldsymbol{q} = \boldsymbol{k}_s - \boldsymbol{k}_i$; its instantaneous intensity is defined by the real part of the Fourier Transform of the particle positions

$$I(\boldsymbol{q}, t) = \sum_{i,j}^{N} \cos[\boldsymbol{q} \cdot \{\boldsymbol{r}_i(t) - \boldsymbol{r}_j(t)\}]. \tag{2.1}$$

The structure factor for a state point in the phase diagram, defined by

$$S(\boldsymbol{q}) = \frac{1}{N} \langle I(\boldsymbol{q}, t) \rangle, \tag{2.2}$$

was calculated from data accumulated over the last 1,000 frames of the simulation. For every frame, the positions of the particles were used to make an image in pixels that was then Fourier transformed. The multiple frame structure factors are then averaged together to produce the final structure factor.

The normalized aperture cross correlation function of Ackerson and Clark is defined by [10]:

$$C(\boldsymbol{k}, \boldsymbol{q}) = \frac{\langle I(\boldsymbol{k}) I(\boldsymbol{k}+\boldsymbol{q}) \rangle}{\langle I(\boldsymbol{k}) \rangle \langle I(\boldsymbol{k}+\boldsymbol{q}) \rangle}, \tag{2.3}$$

where $I(\boldsymbol{k})$ and $I(\boldsymbol{q})$ are the instantaneous intensities at two detectors. The brackets indicate both a time and spatial average; for each time step of the simulation, the entire system is sampled by c0omputing the aperture cross correlation for a number of fixed size apertures covering the spatial extent of the system. By fixing $\boldsymbol{k}$ and putting $|\boldsymbol{k}| = |\boldsymbol{q}|$, we can measure a one-dimensional slice of the full four-dimensional function that depends only on the angle between $\boldsymbol{k}$ and $\boldsymbol{q}$,

$$C(\phi) = \frac{\langle I(\phi_0) I(\phi_0+\phi) \rangle}{\langle I(\phi_0) \rangle \langle I(\phi_0+\phi) \rangle}, \tag{2.4}$$

where $\phi_0$ is a constant.

We have calculated the ACCF by creating diffraction patterns from randomly located, circular apertures across the simulation box. The circular aperture radius is set to four particle diameters. With respect to the aperture size chosen, the calculations reported by Sheu and Rice [38] show that the ACCFs calculated for apertures of three and five particle diameters are sensibly the same (see Fig. 10 of Ref. 38). A number of sample regions were taken from the last 200 frames of the simulation, with the constraint that the total sample area is at least the area of the simulation box. These regions may overlap but were selected such that the positions of their centers are at least one aperture radius away from the edge of the simulation box. To calculate the ACCF of a region, the instantaneous intensity of the original image is cross-correlated with the instantaneous intensity of its rotated image at every integer degree. The ACCF was calculated using data from the last 1,000 frames of the simulation, then averaged over the samples.

## 3. Results

We now examine the character of liquid-phase transient ordered fluctuations as a function of density along the isotherms $T = 0.12$, $T = 0.25$ and $T = 0.30$, and as a function of temperature along the constant density line $\rho = 0.700$. These isotherms and the isochore were selected to sample regions that include the most obvious phase diagram features (see Fig. 2), namely two ordered hexagonal phases in domains separated by a liquid phase and a square ordered phase. The ordered phases have lattice spacings that map respectively to the two length scales of the potential, $\sigma$ and $\sigma_1$. From analyses of the low temperature isotherms ($T < 0.3$) and the corresponding translation and orientation order correlation functions, Ryzhov and coworkers established that in the low-density portion of the phase diagram the liquid and ordered hexagonal solid are separated by continuous transformations to and from an intermediate hexatic phase. However, in the high-density portion of the phase diagram, the transitions between the ordered square solid and the liquid and the transition between the ordered hexagonal and square solids are first order. A comparable study of the high temperature isotherms ($T < 0.3$) reveals only a first order transition from the liquid to the ordered hexagonal solid. We remind the reader that the potential (1.1) also supports a quasi-crystalline phase with 12-fold symmetry [37] that, at low temperature, occupies a very narrow domain between the square and triangular phases, and which we ignore.

### 3.1  $0.38 \leq \varrho \leq 0.46$ along the isotherm $T = 0.12$

The local bond orientation correlation function

$$g_6(r) = \langle \psi_6^*(0)\psi_6(r) \rangle , \qquad (3.1)$$

is a measure of the importance of 6-fold orientation order in the 2D system. In (3.1),

$$\psi_6 = \frac{1}{NN}\Sigma_j \, exp\left[6i\theta(r_{ij})\right] . \qquad (3.2).$$

Here, the index $j$ runs over the *NN* nearest neighbors of the *i*-th particle, found via a Voronoi tessellation, and $\theta(r_{ij})$ is the angle between the straight line (bond) connecting the centers of particles *i* and *j* and an arbitrary reference axis that is held fixed throughout the entire calculation. KTHNY theory predicts that in the hexatic phase $g_6(r) = \langle \psi_6^*(0)\psi_6(r)\rangle \propto r^{-\eta}$ with $0 \leq \eta \leq \frac{1}{4}$, and that $\eta = 1/4$ on the boundary between the hexatic and liquid phases [35]. Using that criterion, Ryzhov and coworkers located the low-density boundary between the hexatic and liquid phases on the $T = 0.12$ isotherm at $\rho = 0.4325$. We display in Fig. 3 the ACCFs and $g_6(r)$ for the density range $0.38 \leq \rho \leq 0.46$. We find that transient ordered fluctuations with 6-fold symmetry are present in the liquid as much as 11% below the hexatic-to-liquid transition density. We find no evidence for ordered fluctuations with any other symmetry. Recalling that the hexatic phase has quasi-long-range orientation order without translation order, as expected, the six-fold symmetry is stronger in the hexatic phase than in the liquid phase. Given the continuous character of the hexatic-to-liquid transition it is not surprising that the intensity of the six-fold symmetry increases continuously over the density range $0.38 \leq \rho \leq 0.46$, as shown in Fig. 3d. It is noteworthy that the full width at half height of the peaks in the ACCF do not change over the density range $0.38 \leq \rho \leq 0.46$ that brackets the liquid-to-hexatic transition density, implying that the distribution of angular ordering of the fluctuations in the liquid domain are the same as in the hexatic domain.

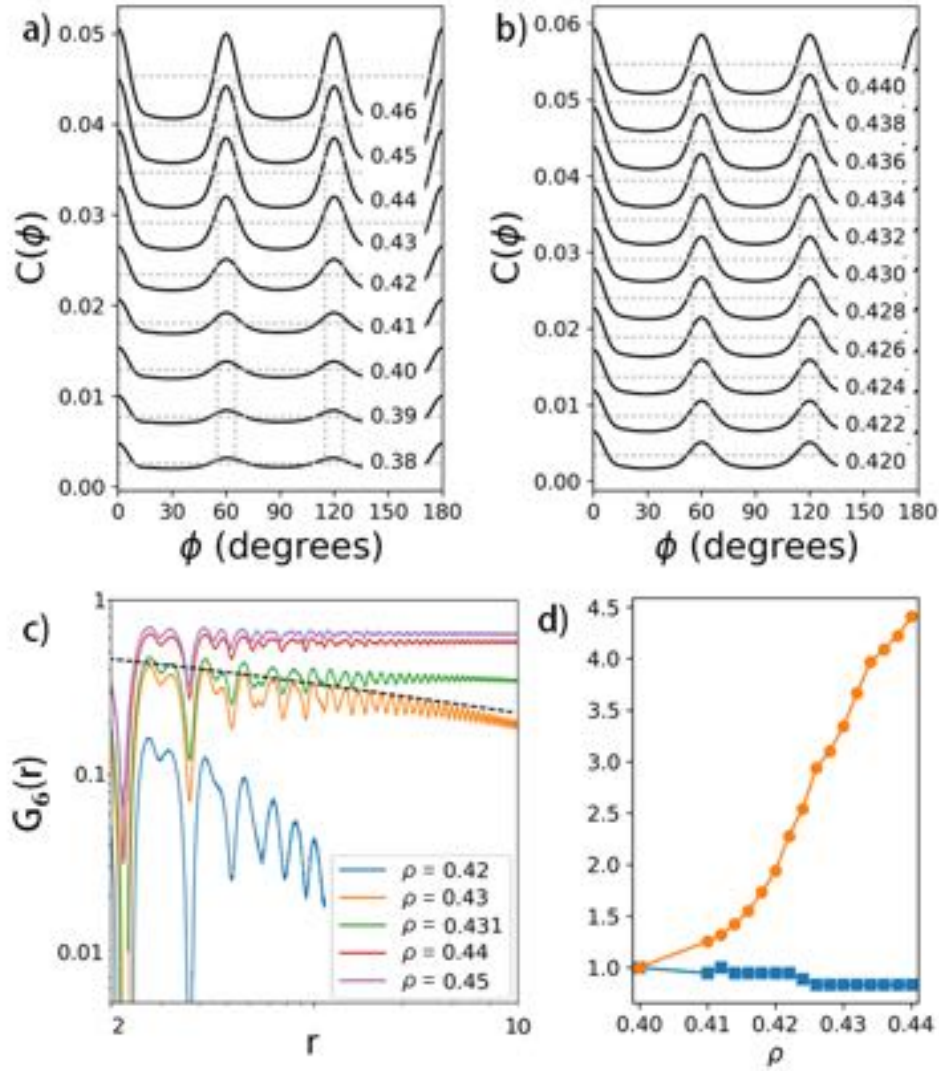

**Figure 3.** (a) ACCFs for T = 0.12 and densities $0.38 \leq \rho \leq 0.46$. (b) ACCFs for T = 0.12 and densities $0.42 \leq \rho \leq 0.44$. (c) $g_6(r)$ for T = 0.12 and densities $0.42 \leq \rho \leq 0.45$. (d) Amplitude (filled circles) and full width at half height (filled squares) of the ACCF peak as a function of density crossing the liquid-to-hexatic transition line, normalized to the peak parameters for $\rho = 0.40$.

### 3.2  $0.58 \leq \rho \leq 0.64$ along the isotherm $T = 0.12$

Staying on the isotherm $T = 0.12$, the stable phase for the domain $0.46 \leq \rho < 0.58$ is ordered hexagonal, for $0.58 \leq \rho \leq 0.60$ the stable phase is hexatic and when $\rho > 0.60$ the system is in a liquid phase. Figure 4 displays the $g_6(r)$ and the ACCFs for this temperature and range of densities. As expected, we find that 6-fold symmetry is present throughout the hexatic domain and that the symmetry persists into the liquid phase above the hexatic-to-liquid transition density. The intensity of the 6-fold symmetry is markedly greater in the hexatic phase than in the liquid phase. Complementary to the behaviour observed on crossing the liquid-to-hexatic transition line $\varrho = 0.43$, the intensity of fluctuations with hexagonal symmetry decreases continuously across the hexatic-to-liquid transition line at $\varrho = 0.596$. On crossing the transition line into the liquid phase the width of the peaks in the ACCF increase somewhat.

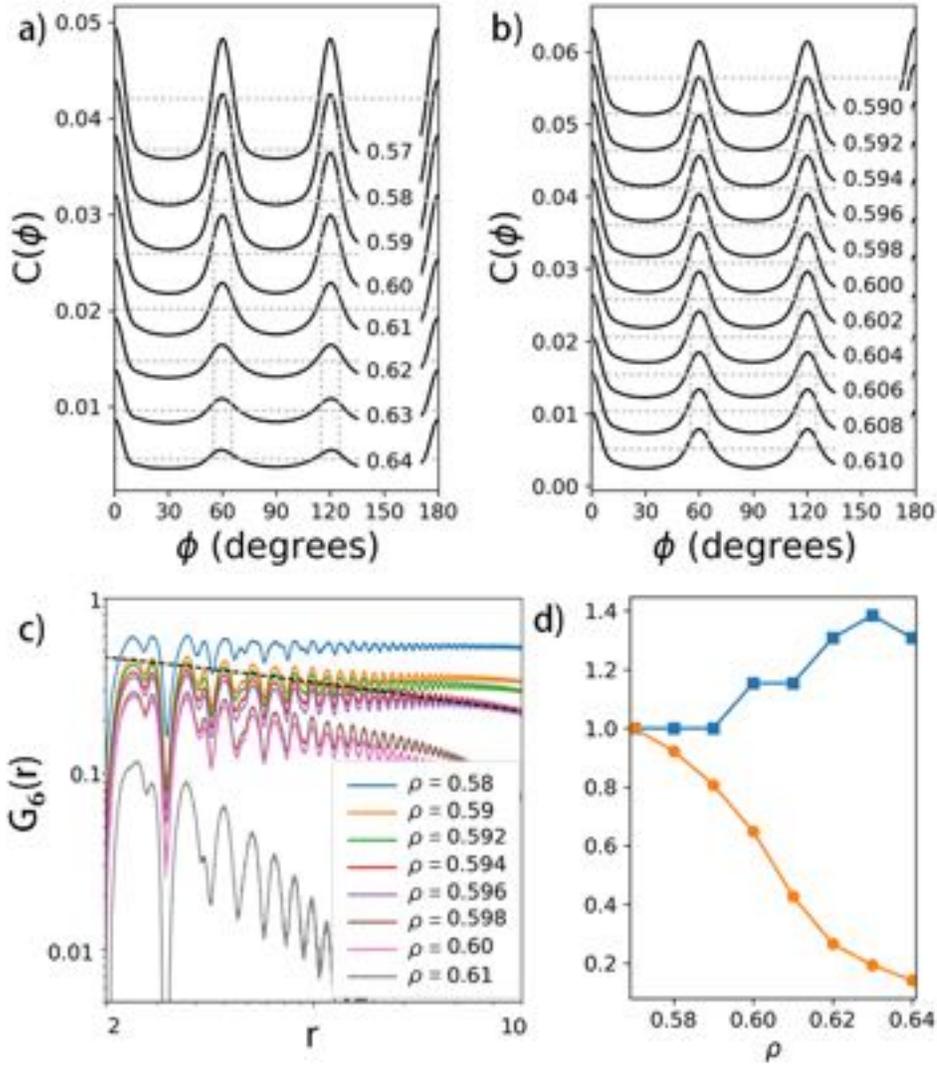

**Figure 4.** (a) ACCFs for $T = 0.12$ and densities $0.57 \leq \rho \leq 0.64$. (b) ACCFs for $T = 0.12$ and densities $0.59 \leq \rho \leq 0.61$. (c) $g_6(r)$ for $T = 0.12$ and densities $0.58 \leq \rho \leq 0.61$. (d) Amplitude (filled circles) and full width at half height (filled squares) of the ACCF peak as a function of density crossing the liquid-to-hexatic transition line, normalized to the peak parameters for $\rho = 0.64$.

### 3.3  $0.67 \leq \rho \leq 0.77$ along the isotherm $T = 0.12$ and $T = 0.12, 0.20, 0.30$ and $0.40$ along the isochore $\rho = 0.70$

The region $0.67 \leq \rho \leq 0.77$ along the isotherm $T = 0.12$ is occupied by the liquid phase, and the structure factor of the system suggests the coexistence region for the transition to the square phase starts around $\rho = 0.770$ (Fig. 5). At this density, azimuthal intensity maxima superposed on a ring of intensity can be distinguished. Given that the ordered solids at the low density and high density ends of this density range have different symmetries, we ask how the structured fluctuations in the liquid between these ordered solid structures change as the density is varied from 0.65 to 0.77.

We display in Figure 6 the ACCFs for several densities in the stable liquid domain and two densities in the stable square solid domain. At the low-density end of this density range, the dominant structured fluctuations are hexagonal. At the upper end of this density range, the dominant structured fluctuations are square. However, both structured fluctuations occur, with varying concentrations, at all densities in the range $0.67 \le \rho \le 0.77$. For example, we show in Figure 7 the ACCFs for four temperatures along the constant density line $\rho = 0.70$, and as demonstrated in Figure 8 the ACCF line-shape is well represented by the sum of contributions from 4-fold and 6-fold ordered fluctuations. It is immediately evident that square and hexagonal ordered fluctuations are simultaneously present, and that the ratio of concentrations of square to hexagonal ordered fluctuations decreases as the temperature increases.

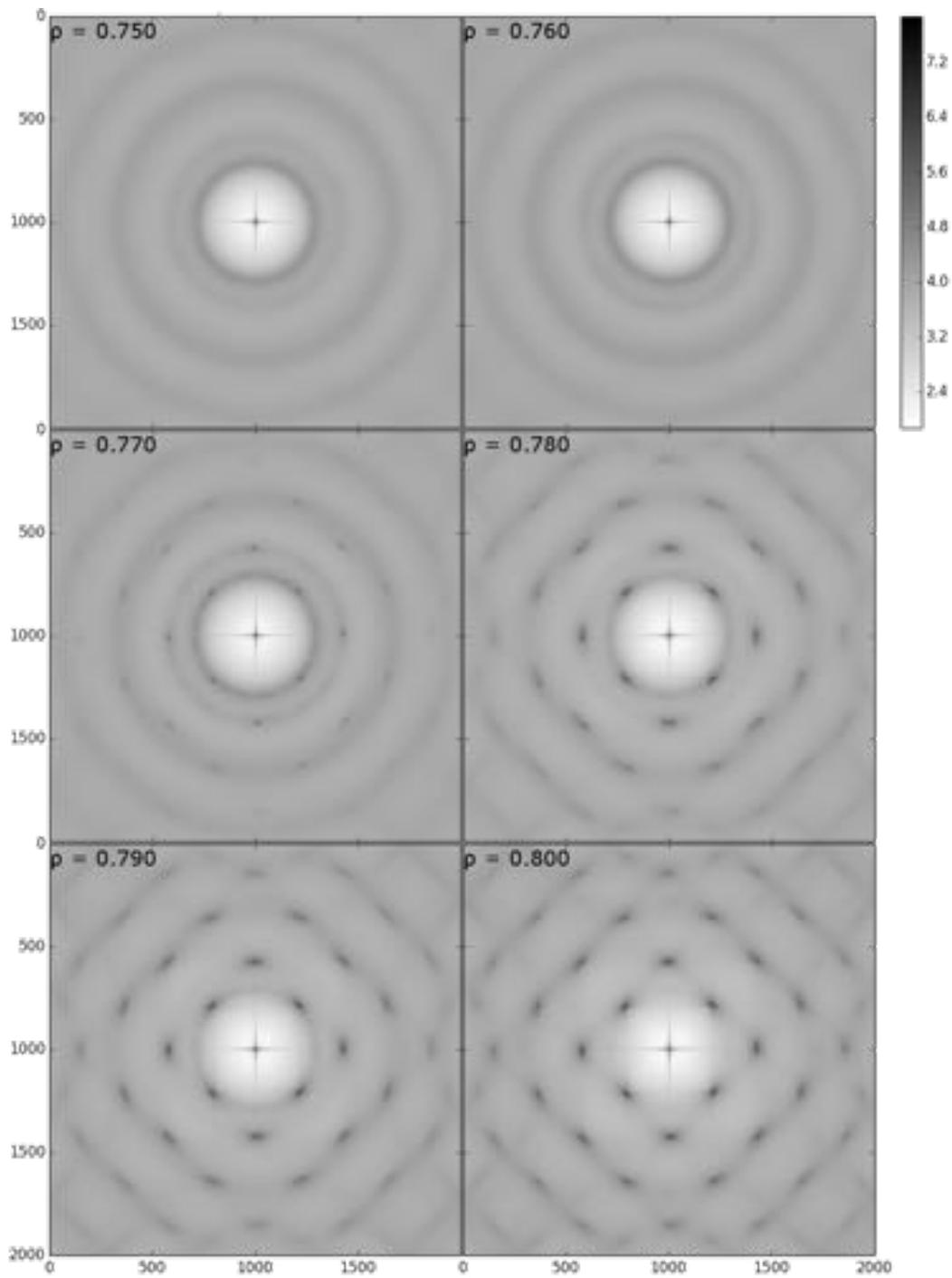

**Figure 5.** Structure functions calculated along the $T = 0.12$ isotherm. For $0.750 \leq \rho \leq 0.800$ the structure functions imply coexistence of the liquid and square solid phases.

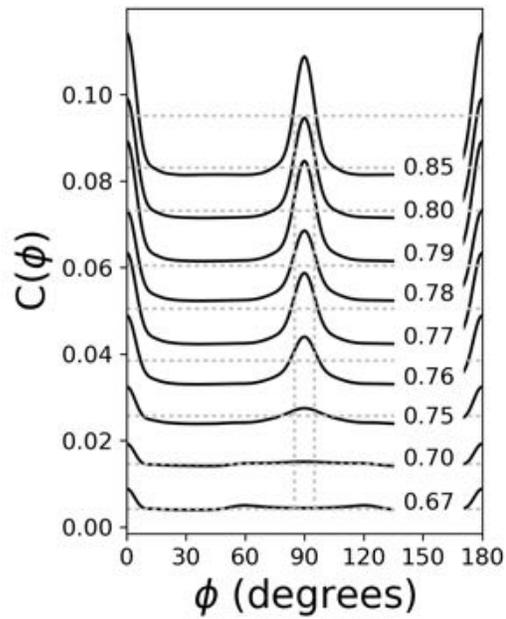

**Figure 6.** Several ACCFs at densities in the stable liquid domain $0.67 \leq \rho \leq 0.76$ and in the stable square solid domain $0.80 \leq \rho \leq 0.85$. The ACCF for $0.77 \leq \rho \leq 0.79$, as implied by the structure function displayed in Figure 5c, shows the fluctuations in the liquid-square solid coexistence domain. The temperature is $T = 0.12$.

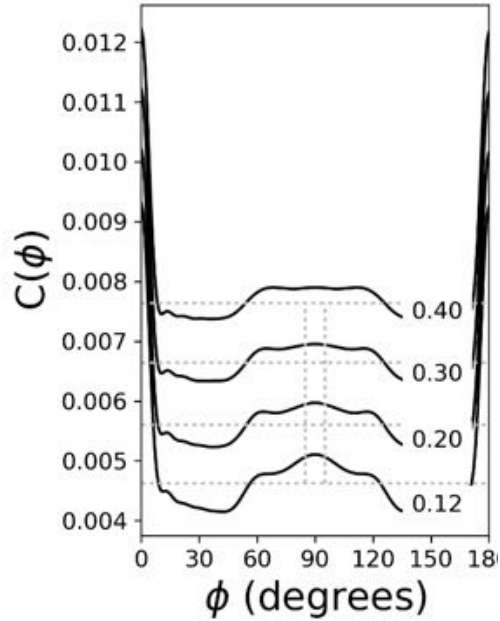

**Figure 7.** ACCFs along the isochore $\rho = 0.70$ at $T = 0.12, 0.20, 0.30,$ and $0.40$. The ACCFs for $T = 0.20, 0.30$ and $0.40$ are translated up by successive increments of $0.1$.

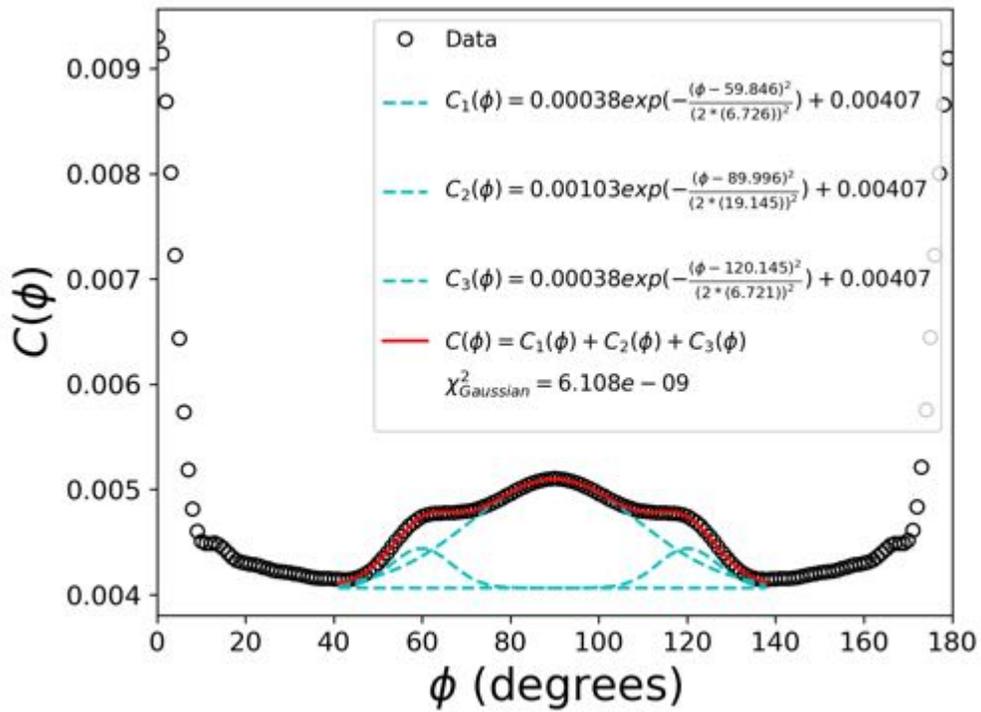

**Figure 8.** The fit of amplitudes of simultaneous square and hexagonal ordered fluctuations to the ACCF line-shape for $\varrho = 0.70$ at $T = 0.12$.

**3.4 $0.40 \leq \varrho \leq 0.90$ along the isotherm $T = 0.25$ and $0.40 \leq \varrho \leq 0.95$ along the isotherm $T = 0.30$**

We now examine the structures of the fluctuations along the isotherms $T = 0.25$ and $T = 0.30$ for the density ranges $0.40 \leq \varrho \leq 0.90$ and $0.40 \leq \varrho \leq 0.95$ respectively. The $T = 0.25$ isotherm comes very close to the maximum temperature of the transition line for the higher density liquid-to-square transition, but for $\varrho \leq 0.90$ both isotherms lie entirely in the liquid domain. When $T = 0.30$, the onset of the liquid-to-ordered-hexagonal-solid is at about $\varrho = 0.92$ (see Fig. 2). Indeed, the structure functions calculated at sample densities along the $T = 0.30$ isotherm, shown in Figure 9, display the characteristic continuous ring structure of the disordered liquid. However, the ACCFs for those sample densities when $T = 0.25$ and $T = 0.30$, shown in Figures 10a and 10b, reveal the existence of structured fluctuations that can be correlated with the phase transitions that lie at lower temperature. Thus, at the lowest sample density shown in Figures 8 and 9, namely $\varrho = 0.40$, the phase point is far from the liquid-to-hexatic transition line, and the amplitudes of the fluctuations are quite small, with weak hints that both hexagonal and square transient structures are populated. When the density is $\varrho = 0.50$, the phase point is much closer to the liquid-to-hexatic transition line, the amplitude of the hexagonal fluctuations is much greater than when $\varrho = 0.40$, and the amplitude of the square fluctuations is greatly diminished, noting that the amplitude of the trough between the ACCF peaks at 60° and 120° is not as small as that between the peaks at 120° and 180°. When $\varrho = 0.60$ the overall ACCF structure is very similar to that when $\varrho = 0.50$. When $\varrho = 0.70$ the phase point is about halfway between the critical densities for the liquid-hexagonal solid and liquid-square solid transitions, and the amplitudes of the hexagonal and square structure fluctuations are comparable (see Section 3.3). At higher densities, $\rho = 0.80$ and $\rho = 0.90$ the hexagonal fluctuations appear to be favoured relative to the square fluctuations.

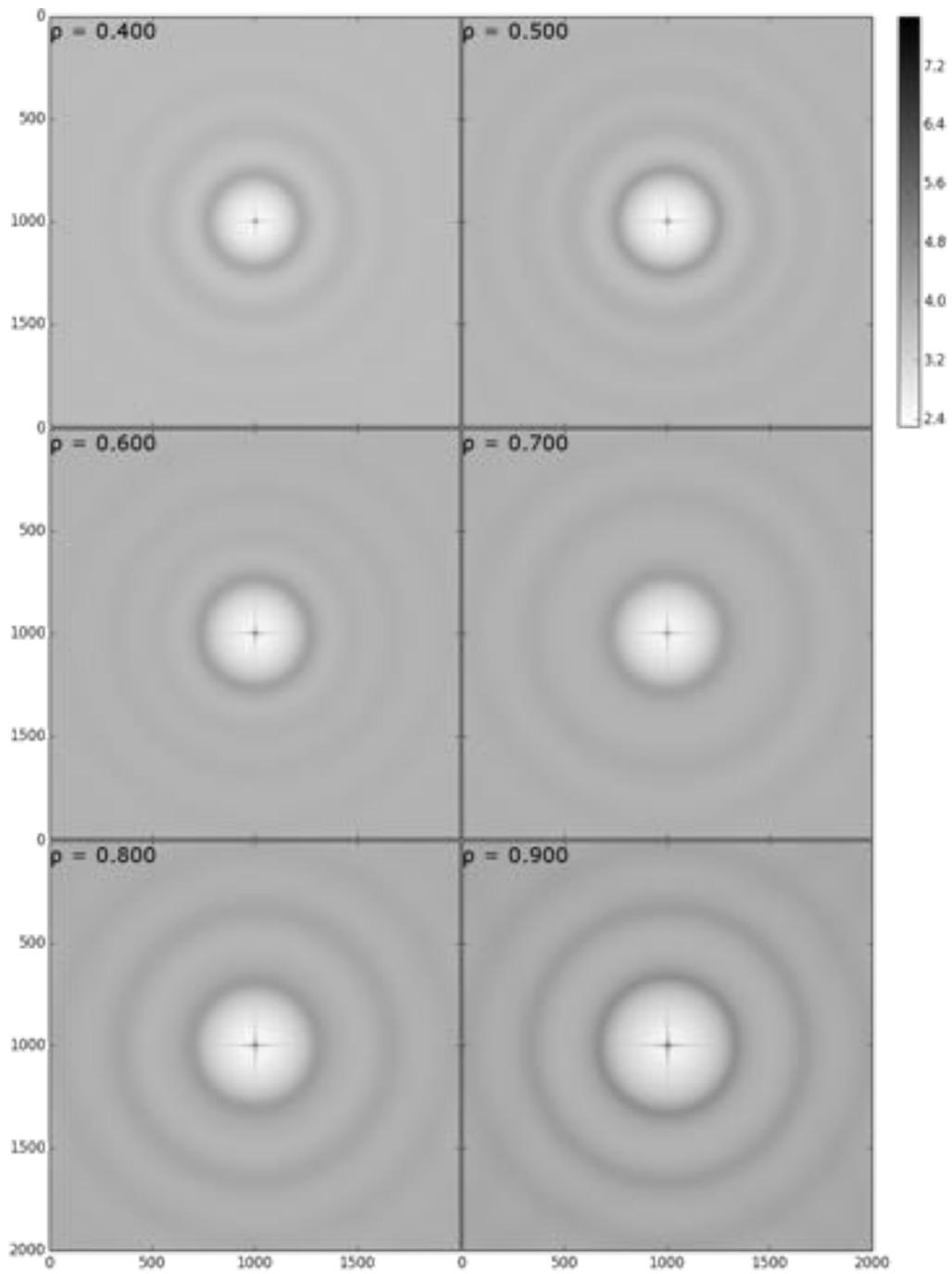

**Figure 9.** Structure functions calculated along the $T = 0.30$ isotherm imply the existence of only the liquid phase.

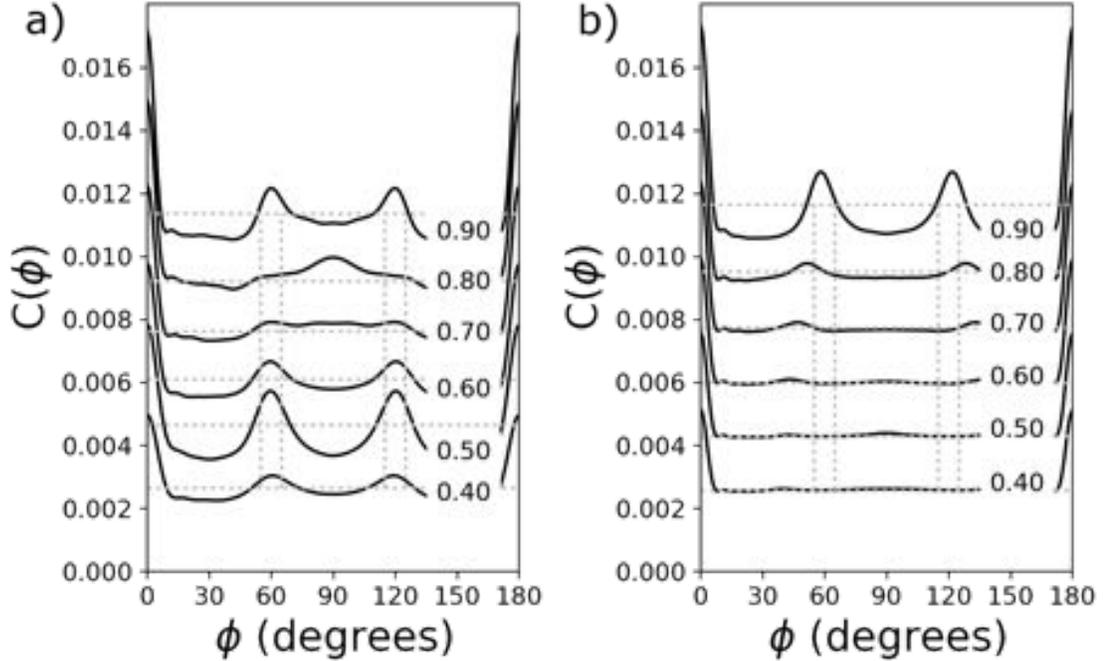

**Figure 10.** ACCFs for several densities along (a) the isotherm $T = 0.25$ and (b) the isotherm $T = 0.30$.

## 4. Discussion

    We have used the ACCF of a 2D liquid to obtain information about the structure of transient fluctuations that cannot be obtained from other characterizations of the liquid. We emphasize this point with a composite graphic illustration of the information content of several characterizations of thermodynamic states of the 2D Rhyzov liquid studied in this paper, shown in Fig. 11. The top row of panels in Fig. 13 displays the Voronoi tessellations of configurations of particles in the liquid state for three densities, $\rho = 0.420, 0.700$ and $0.750$, all at the reduced temperature $T = 0.12$. In each of these states the Voronoi tessellation clearly displays disorder in the particle configuration via the presence of a dense distribution of cell sizes. The numbers of Voronoi cells in each state with less than and more than six sides, shown in the second row of panels, are large, and the corresponding diffraction patterns, shown in the third row of panels, are continuous rings with uniform azimuthal intensity. None of these representations provides a signature of the structured fluctuations in the liquid that are revealed by the ACCFs shown in the bottom row of panels.

    The results of the simulations reported in this paper extend the findings of Sheu and Rice [38]. They showed that ordered fluctuations in a q2D hard sphere system confined between smooth hard walls exist well into the stable liquid region, that the character of the fluctuations is sensitive to the separation of the confining boundaries applied to the system, and that the fluctuations exhibited a very limited range of structural motifs. In that q2D system the structures of the stable ordered phases follow the sequence $1\Delta \rightarrow 2\square \rightarrow 2\Delta \rightarrow 3\square \rightarrow \cdots$ as the number of

layers between the walls increases from 1 to 2 to 3 … . Furthermore, the ordered fluctuations assume the same symmetry, either hexagonal or square, as the solid to which the liquid freezes at densities about 2% below freezing, while deeper in the liquid phase they assume the symmetry

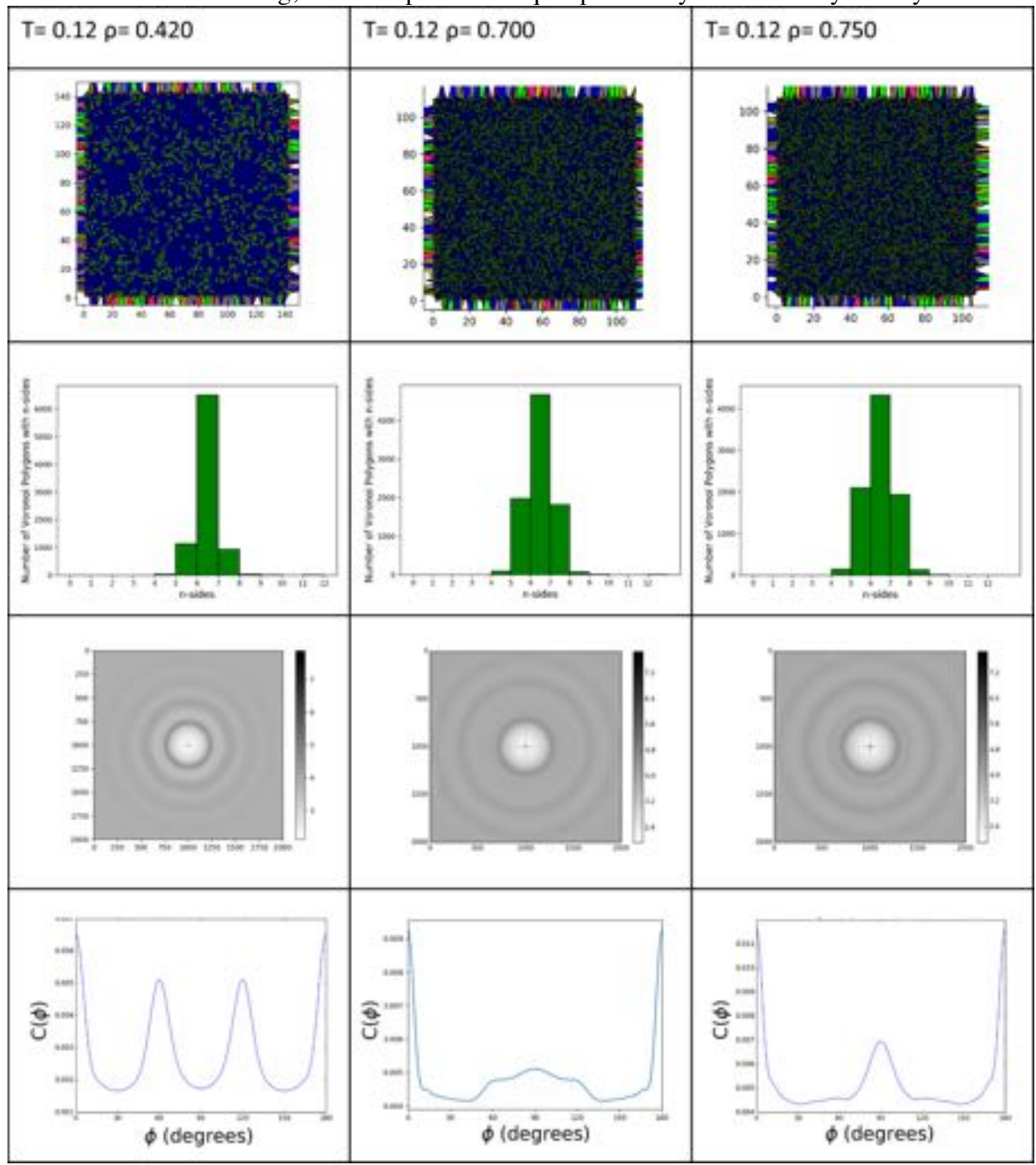

**Figure 11.** Comparison of characterizations of structure in the 2D Rhyzov liquid. The top panel displays Voronoi tessellations of particle configurations with densities $\rho = 0.420, 0.700$ and $0.750$. The Voronoi cells are labelled by the number of sides: $n < 3$ (bright pink), $n = 3$ (teal), $n = 4$ (red), $n = 5$ (yellow), $n = 6$ (blue), $n = 7$ (neon green), $n > 7$ (violet). The second row of panels displays histograms of the numbers of Voronoi cells with n sides, the third row

of panels displays the diffraction pattern of the liquid and the bottom row of panels displays the ACCFs of the liquid.

of the solid that is stable at slightly smaller plate separation. The fluctuations found do not exhibit other ordered structures. The Ryzhov pair potential, which supports a richer set of 2D ordered phases than does the hard sphere potential, also supports only fluctuations with motifs related to those stable solid structures. We regard these findings, that there is no evidence for structured fluctuations in the liquid phase with other symmetries than those of the ordered phases of the system, to be the most striking result of our calculations. Whether this finding applies generally to fluctuations in systems with other interaction potentials remains to be established. Indeed, the relationship between the structures of transient fluctuations in a liquid and the particle-particle pair potential is largely unexplored.

We argue that the ACCF provides a unique tool for exploring the structures of fluctuations in the liquid phase. It is not obvious how to guess what these structures will be from knowledge of the pair potential and/or the ordered phases that the potential supports. For example, it is now known [4,5] that a central pair potential can support the formation of 2D honeycomb and Kagome lattices, both of which have empty lattice sites. The ACCFs of these systems can provide information concerning the structures of transient fluctuations in the liquids prior to transition to the ordered states. In particular, we are interested to learn if the fluctuations in the mobile liquids exhibit structures that mimic the local structures in the solids, or if they have a conventional hexagonal motif that "opens" on solidification. The ACCF of a system can also provide information concerning how internal constraints affect local structure in the liquid. For example, for a system very like that studied in this paper but with a longer soft-shoulder range, Ryzhov and coworkers [42] showed that the phase diagram becomes more complex; the phase diagram has a lower density that has a melting transition with two continuous steps consistent with KTHNY theory and a higher density domain with a first order solid-to-liquid transition with no intermediate hexatic phase. In this system, adding quenched disorder by randomly pinning particles at 0.1% concentration causes the high density melting scenario to split into a continuous solid-to-hexatic transition plus a first order hexatic-to-liquid transition. This pinning procedure also significantly broadens the stability range of the hexatic phase. A study of the ACCF of this system should reveal what happens to the fluctuations in the liquid when the system is partially pinned, and thereby illuminate the basis for the shift in phase boundaries generated by the pinning. These matters, currently under investigation, will be discussed in a future paper.

## 5. Acknowledgments

We thank Prof. V. N. Ryzhov and coworkers for permission to republish Figure 1 and Figure 2 and Alex Smith, Ryan Zarcone, and Dan Older for their programming assistance and advice. The research reported in this paper was primarily supported by the University of Chicago Materials Research Science and Engineering Center, funded by National Science Foundation Grant No. DMR-1420709, and partially supported by a Senior Mentor Grant from the Camille and Henry Dreyfus Foundation (Grant No. SI-14-014). B.L. acknowledges support from Chem. Mat-CARS (NSF/CHE-1346572). L.N. thanks the Jeff Metcalf Internship Program for their support.